\documentclass[11pt,showpacs,aps,nofootinbib]{revtex4}
\usepackage{amsmath,amssymb,graphicx,epsfig,color,xcolor,ulem,cancel,datetime}
\def\comment#1{}

\def\nn{\nonumber}
\def\dst{\mathrm{dS_{2}}}
\def\dsf{\mathrm{dS_{4}}}
\def\x{\mathrm{x}}

\def\v{\mathrm{in}}
\def\e{\mathrm{eff}}
\begin{document}
%%%%%%%%%%%%%%%%%%%%%%%%%%%%%%%%%%%%%%%%%%%%%%%%%%%%%%%%%%%%%%%%%%%%%%%%%%%%%%%%%%%%%%%%%%%%%%%%%%%%%%%%%%%%%%%%%%%%%%%%%%%%%%%%%%%%%%%%%%%%%%%%%%%%%%%%%%%%%%%
%%%%%%%%%%%%%%%%%%%%%%%%%%%%%%%%%%%%%%%%%%%%%%%%%%%%%%%%%%%%%%%%%%%%%%%%%%%%%%%%%%%%%%%%%%%%%%%%%%%%%%%%%%%%%%%%%%%%%%%%%%%%%%%%%%%%%%%%%%%%%%%%%%%%%%%%%%%%%%%
\title{Energy-momentum tensor and effective Lagrangian of scalar QED with a nonminimal coupling in 2D de~Sitter spacetime}
\author{Marzieh~Akbari~Ahmadmahmoudi}%\email{marzyehakbari088@gmail.com}
\author{Ehsan~Bavarsad}\email{bavarsad@kashanu.ac.ir}
\affiliation{Department of Physics, University of Kashan, 8731753153, Kashan, Iran}
%\date{\red{\tt{Revised Version,~\today~at~\currenttime}}}
%%%%%%%%%%%%%%%%%%%%%%%%%%%%%%%%%%%%%%%%%%%%%%%%%%%%%%%%%%%%%%%%%%%%%%%%%%%%%%%%%%%%%%%%%%%%%%%%%%%%%%%%%%%%%%%%%%%%%%%%%%%%%%%%%%%%%%%%%%%%%%%%%%%%%%%%%%%%%%%
%%%%%%%%%%%%%%%%%%%%%%%%%%%%%%%%%%%%%%%%%%%%%%%%%%%%%%%%%%%%%%%%%%%%%%%%%%%%%%%%%%%%%%%%%%%%%%%%%%%%%%%%%%%%%%%%%%%%%%%%%%%%%%%%%%%%%%%%%%%%%%%%%%%%%%%%%%%%%%%
\begin{abstract}
We have studied the induced one-loop energy-momentum tensor of a massive complex scalar field within the framework of nonperturbative quantum electrodynamics (QED) with a uniform electric field background on the Poincar\'e patch of the two-dimensional de~Sitter spacetime ($\dst$). We also consider a direct coupling the scalar field to the Ricci scalar curvature which is parameterized by an arbitrary dimensionless nonminimal coupling constant. We evaluate the trace anomaly of the induced energy-momentum tensor. We show that our results for the induced energy-momentum tensor in the
zero electric field case, and the trace anomaly are in agreement with the existing literature. Furthermore, we construct the one-loop effective Lagrangian from the induced energy-momentum tensor.
\end{abstract}
\pacs{04.62.+v,11.10.Kk,11.10.Gh}
\maketitle
%\tableofcontents
%%%%%%%%%%%%%%%%%%%%%%%%%%%%%%%%%%%%%%%%%%%%%%%%%%%%%%%%%%%%%%%%%%%%%%%%%%%%%%%%%%%%%%%%%%%%%%%%%%%%%%%%%%%%%%%%%%%%%%%%%%%%%%%%%%%%%%%%%%%%%%%%%%%%%%%%%%%%%%%
%%%%%%%%%%%%%%%%%%%%%%%%%%%%%%%%%%%%%%%%%%%%%%%%%%%%%%%%%%%%%%%%%%%%%%%%%%%%%%%%%%%%%%%%%%%%%%%%%%%%%%%%%%%%%%%%%%%%%%%%%%%%%%%%%%%%%%%%%%%%%%%%%%%%%%%%%%%%%%%
%\iffalse
%\section*{Main Points:}
%\begin{itemize}
%\item
%\end{itemize}
%\fi
%%%%%%%%%%%%%%%%%%%%%%%%%%%%%%%%%%%%%%%%%%%%%%%%%%%%%%%%%%%%%%%%%%%%%%%%%%%%%%%%%%%%%%%%%%%%%%%%%%%%%%%%%%%%%%%%%%%%%%%%%%%%%%%%%%%%%%%%%%%%%%%%%%%%%%%%%%%%%%%
%%%%%%%%%%%%%%%%%%%%%%%%%%%%%%%%%%%%%%%%%%%%%%%%%%%%%%%%%%%%%%%%%%%%%%%%%%%%%%%%%%%%%%%%%%%%%%%%%%%%%%%%%%%%%%%%%%%%%%%%%%%%%%%%%%%%%%%%%%%%%%%%%%%%%%%%%%%%%%%
\section{\label{sec:intro}introduction}
The basic framework of quantum field theory in curved spacetime was originally introduced by Parker \cite{Parker:1969au} in the late of 1960s and followed by others. In the Parker's pioneer work quantization of fields was described and the average density of created particles in an expanding universe was derived. Progress was also being made on this issue, when in the mid of 1970s Hawking discovered \cite{Hawking:1974sw} that a black hole emits as a
blackbody due to the particle creation which is known as Hawking radiation. With these discoveries the development of quantum field theory in curved spacetime received much further motivations; see, e.g., \cite{Birrell:1982ix,Parker:2009uva,Wald:1995yp} for introduction. Indeed, a general curved spacetime is not invariant under transformations of the Poincar\'e group, as a consequence there is no a natural set of field modes which are invariant under Poincar\'e transformations. This ambiguity leads to ambiguity in definition of particle concept. The fact that the field modes are defined on the whole of at least a large patch of spacetime illustrates the global nature of the particle concept. This is in contrast with the at least quasi-local
nature of physical detectors. Hence, it is more advantageous to construct locally-defined quantities; see \cite{Birrell:1982ix} for a comprehensive
review. One such object of interest is the energy-momentum tensor which is constructed from fields and their derivatives at the same point of spacetime. There are two important reasons for studying energy-momentum tensor \cite{Birrell:1982ix,Parker:2009uva}. In Einstein's equation the energy-momentum
tensor appears as a source term of the gravitational field, hence it can be used to investigate the backreaction effects of the matter on the dynamics of gravitational field. Also, it is a useful quantity to explore the physical properties of the quantum fields. Thus, studying the energy-momentum tensor of quantized fields get more interesting when the cosmological spacetimes have been considered.
\par
The regularized and renormalized energy-momentum tensor for different quantum fields in curved spacetime has been extensively studying using various methods. The renormalized energy-momentum tensor of a quantized neutral scalar field propagating in a spacetime of the type of Friedmann-Lemaitre-Robertson-Walker (FLRW) universes has been analyzed in several cases of interest: (1) For a massive field with the arbitrary \cite{Bunch:1980vc,Bunch:1978yq,Habib:1999cs,LopezNacir:2007fdi,LopezNacir:2007wvc,Zhang:2019urk,Anderson:1987yt}, minimal \cite{Fulling:1974zr,Parker:1974qw}, and conformal coupling \cite{Anderson:2013ila,Anderson:2013zia,Fulling:1974pu} to the Ricci scalar curvature. (2) For
a massless field with an arbitrary \cite{Zhang:2019gtg,Bunch:1978yw}, and conformal \cite{Davies:1977ze} coupling to the Ricci scalar curvature. In \cite{Dowker:1975tf}, the energy-momentum tensor and effective Lagrangian of a massive neutral scalar field with the nonminimal coupling to the Ricci
scalar curvature in a de~Sitter spacetime (dS) have been calculated by using the dimensional regularization. In Ref.~\cite{Mottola:1984ar}, to study the effects of the particle creation in a dS the finite energy-momentum tensor of a massive neutral scalar field with the nonminimal coupling to the Ricci scalar curvature was evaluated by computing the difference in energy-momentum between the in and out-vacuum states. Then, it was realized that the energy-momentum tensor of the created particles describes a perfect fluid with vacuum equation of state which vanishes for massless, conformally coupled field. Also, the author discovered that the invariant vacuum state and the effective cosmological constat decay due to the particle creation. In the work
of \cite{Markkanen:2016aes} the energy-momentum tensor of a quantum noninteracting, massive, and nonminimally coupled scalar field in a dS has been investigated. And, it was shown as a consequences of the quantum backreaction effects that there may exist a phase of superacceleration in which the
Hubble constant amplifies. With the aim of developing the adiabatic expansion for the case of fermion fields, the average number of created particles and regularized energy-momentum tensor of a noninteracting, massive Dirac field in a spatially flat FLRW universe have been computed in Refs.~\cite{Landete:2013axa,Landete:2013lpa,delRio:2014cha,Ghosh:2015mva,Ghosh:2016epo}.
\par
In order to make one step forward in the context of particle creation in curved spacetime, it seems natural to add an electromagnetic gauge field interacting with the quantum matter field. In fact a strong electromagnetic field background in the Minkowski spacetime can create pairs of particles \cite{Schwinger:1951nm,Sauter:1931zz,Heisenberg:1935qt} which is known as the Schwinger effect; see \cite{Gelis:2015kya,Dunne:2004nc} for a comprehensive review. Indeed, the physical mechanism underlying the Schwinger effect is analogous to that of the gravitational particle creation phenomena in curved spacetime \cite{Martin:2007bw}. Thus, studying the Schwinger effect in the cosmological spacetimes would be interesting because it may amplify the gravitational particle creation process; see \cite{Kim:2019joy} and references therein for a review. It is a well accepted paradigm \cite{Durrer:2013pga} that strong electric and magnetic fields might be generated in the early universe which motivates the study of Schwinger effect in the dS. Investigation
of the Schwinger effect in the dS was initiated by \cite{Garriga:1993fh,Garriga:1994bm}. The Schwinger effect and the rate of scalar pair creation process in the presence of a uniform electric field background have been analyzed in a dS of two \cite{Garriga:1994bm,Frob:2014zka,Cai:2014qba,Kim:2008xv,%
Hamil:2018rvu}, four \cite{Kobayashi:2014zza}, and general \cite{Bavarsad:2016cxh} dimensions, by using the technique of Bogoliubov transformation that requires semiclassical conditions. By using this technique, the influence of a uniform and conserved flux magnetic field on the creation of scalar pairs
by the Schwinger mechanism in a four dimensional de~Sitter spacetime ($\dsf$) has been explored \cite{Bavarsad:2017oyv,Bavarsad:2018lvn}; see also \cite{Moradi:2009zz}. The authors of \cite{Bavarsad:2017oyv,Bavarsad:2018lvn} found that a strong magnetic field can intensify the Gibbons-Hawking
radiation \cite{Gibbons:1977mu} of $\dsf$ even when there is no an electric field. It is worth mentioning that the creation of Dirac pairs by Schwinger mechanism in a dS has also been studied using semiclassical methods \cite{Villalba:1995za,Haouat:2012ik,Stahl:2015gaa,Stahl:2015cra,Haouat:2015uaa}, with the main conclusion that a strong electric field enhances significantly the gravitational pair creation. The Schwinger effect has been explored in a flat FLRW \cite{Haouat:2012dr}, anti-de~Sitter \cite{Cai:2014qba,Kim:2008xv,Pioline:2005pf,Samantray:2016uuj}, and a charged black hole \cite{Chen:2012zn,Kim:2016dmm,Chen:2016caa} spacetimes; also, widely considered in studies relevant to the inflationary universe scenarios \cite{Geng:2017zad,Shakeri:2019mnt,Stahl:2018idd,Tangarife:2017rgl,Kitamoto:2018htg,Giovannini:2018qbq,Sharma:2017ivh,Chua:2018dqh,Sobol:2018djj,%
Sobol:2019xls,Domcke:2019qmm,Gorbar:2019fpj}.
\par
The investigation of the Schwinger pair creation in a dS, by using the Bogoliubov transformation method, needs to define an adiabatic out-vacuum state at
late times in addition to the adiabatic in-vacuum state at early times, which in turn requires to impose semiclassical conditions. In the semiclassical conditions, either the mass of the particle or the eclectic potential energy across the Hubble radius or both must be very larger than the energy scale determined by the curvature of the spacetime \cite{Frob:2014zka,Kobayashi:2014zza}. On the contrary, the in-vacuum state of quantum fields in dS satisfies the adiabatic conditions at all times. Hence, computation of expectation values of physical quantities, such as current and energy-momentum tensor, in the in-vacuum state enables us to probe wider ranges of the related parameters. The regularized in-vacuum expectation value of the current of a charged scalar field, caused by a uniform electric field background, has been computed in two \cite{Frob:2014zka}, three \cite{Bavarsad:2016cxh}, and four \cite{Kobayashi:2014zza} dimensional de~Sitter spacetimes. The authors showed that in the strong electric field regime, the induced current asymptotically approaches the semiclassical current. In particular, it was reported that for an essentially light scalar field in the weak electric field regime, the induced current has an inversely proportional response to the electric field, which is referred to as the infrared hyperconductivity phenomenon \cite{Frob:2014zka,Kobayashi:2014zza,Bavarsad:2016cxh}. The derived results for the induced current in $\dsf$ \cite{Kobayashi:2014zza} have been verified
by applying an alternative regularization that is the point-splitting method \cite{Hayashinaka:2016dnt}, and also calculating the current by using the uniform asymptotic approximation method \cite{Geng:2017zad}. An investigation of the influence of a uniform magnetic field background on the current of created scalar pairs by a parallel uniform electric field background in $\dsf$ illustrates that there is a period of infrared hyperconductivity \cite{Bavarsad:2017oyv,Bavarsad:2018lvn}. In $\dst$ \cite{Stahl:2015gaa} and $\dsf$ \cite{Hayashinaka:2016qqn}, the in-vacuum expectation value of the current of a Dirac field coupled to a uniform electric field background has been analyzed. And the authors come to the conclusion that in the infrared regime the fermionic current is free of the hyperconductivity phenomenon, as opposed to the scalar current. The negative current phenomenon is another remarkable feature of the regularized current in $\dsf$, which is occurred for the scalar fields with essentially small masses
\cite{Kobayashi:2014zza,Hayashinaka:2016dnt} and the Dirac fields with any mass \cite{Hayashinaka:2016qqn} in a certain range of the electric field
strength when the current points in the opposite direction to the electric field background. By introducing a novel condition for renormalization of the in-vacuum expectation values of the scalar and Dirac currents in $\dsf$, it was shown that the infrared hyperconductivity period would be removed from the scalar current, however the negative current phase would still be present \cite{Hayashinaka:2018amz}. A satisfactory explanation for the behaviours of the current has been given in Ref.~\cite{Banyeres:2018aax}.
\par
The aim of this paper is to study the expectation value of the energy-momentum tensor of a massive complex scalar field coupled to a uniform electric
field background in the Poincar\'e patch of $\dst$. We also consider a direct coupling the scalar field to the Ricci scalar curvature of $\dst$ which is parameterized by an arbitrary dimensionless nonminimal coupling constant. To compute the expectation value, we will choose the in-vacuum state of the quantized scalar field, because it is an adiabatic and Hadamard state \cite{Garriga:1994bm,Frob:2014zka}. We evaluate the expectation value to one-loop order, hence the ultraviolet divergences will naturally occur in our calculations. To remove these ultraviolet divergences, we will use the method of adiabatic regularization \cite{Anderson:1987yt,Fulling:1974zr,Parker:1974qw,Fulling:1974pu,Parker:1968mv}, because it is comparatively simpler than the other methods, such as, point-splitting regularization \cite{Bunch:1978yq,Bunch:1978yw,Davies:1977ze,Christensen:1976vb} and dimensional regularization \cite{Dowker:1975tf,Candelas:1975du}. It was verified \cite{Birrell:1978pro} that the adiabatic and point-splitting regularization methods will lead to
the equivalence result in spatially flat FLRW spacetimes. There has been several studies to investigate the energy-momentum tensor of created scalar and Dirac pairs by a uniform electric field in a dS. The energy-momentum tensor of created scalar pairs by a uniform electric field in a dS of general
dimension was calculated by using the Bogoliubov coefficients in the two limiting regimes: the heavy scalar field \cite{Bavarsad:2016cxh}, and the strong electric field \cite{Bavarsad:2017wbe}; which leads to a decay of the Hubble constant. An investigation of the gravitational consequences of scalar pair creation due to a uniform electric field background in the three \cite{Bavarsad:2018jpr} and four \cite{Bavarsad:2019jlg} dimensional dS has been made by calculating the regularized expectation value of the trace of energy-momentum tensor in the in-vacuum state. Recently, in \cite{Botshekananfard:2019zak}
for a massive Dirac field coupled to a uniform electric field background in the Poincar\'e patch of $\dst$, the adiabatic regularized in-vacuum
expectation value of the energy-momentum tensor has been evaluated.
A common conclusion of \cite{Bavarsad:2018jpr,Bavarsad:2019jlg,Botshekananfard:2019zak} was that the sign of the trace can be either positive or negative, depending on the intensities of the parameters mass and electric field. Consequently, the Hubble constant decreases under the condition that the trace is positive, in contrast it increases when the trace is negative. The significant achievement of this paper is the construction of the effective Lagrangian from the regularized energy-momentum tensor.
\par
The paper proceeds as follows. In the next section, we briefly introduce the elements of our analysis. In Sec.~\ref{sec:compu}, the expectation value
of the energy-momentum tensor in the in-vacuum state, and the complete set of appropriate adiabatic counterterms are computed, we then obtain the
regularized energy-momentum tensor. In Sec.~\ref{sec:impli}, the regularized energy-momentum tensor is analyzed, then we use it to derive the trace
anomaly and construct the effective Lagrangian. Eventually, our conclusions are drawn in Sec.~\ref{sec:concl}. In the appendix, we include essential
information which is needed to study of the paper.
%%%%%%%%%%%%%%%%%%%%%%%%%%%%%%%%%%%%%%%%%%%%%%%%%%%%%%%%%%%%%%%%%%%%%%%%%%%%%%%%%%%%%%%%%%%%%%%%%%%%%%%%%%%%%%%%%%%%%%%%%%%%%%%%%%%%%%%%%%%%%%%%%%%%%%%%%%%%%%%
%%%%%%%%%%%%%%%%%%%%%%%%%%%%%%%%%%%%%%%%%%%%%%%%%%%%%%%%%%%%%%%%%%%%%%%%%%%%%%%%%%%%%%%%%%%%%%%%%%%%%%%%%%%%%%%%%%%%%%%%%%%%%%%%%%%%%%%%%%%%%%%%%%%%%%%%%%%%%%%
\section{\label{sec:field}the quantum scalar field in electric and dS backgrounds}
In this section we will introduce the elements of model under consideration and setup our analysis. We imagine a massive charged scalar field which interacts with a uniform electric field background in the Poincar\'e patch of $\dst$. Hence, the scalar field is under the influence of two backgrounds,
i.e., the electromagnetic and gravitational fields which are supposed to be unaffected by the dynamics of the scalar field. The classical action of a complex scalar field $\varphi(x)$ of mass $m$ and electric charge $e$ which is coupled to an electromagnetic gauge field $A_{\mu}$ in the $\dst$ is
\begin{equation}\label{action}
S=\int d^{2}x\sqrt{-g}\Big\{ g^{\mu\nu}\big(\partial_{\mu}+ieA_{\mu}\big)\varphi\big(\partial_{\nu}-ieA_{\nu}\big)\varphi^{\ast}-\big(m^{2}+\xi R\big)\varphi\varphi^{\ast} \Big\},
\end{equation}
where $\xi$ is a dimensionless nonminimal coupling constant and $R=2H^{2}$, written in terms of the Hubble constant $H$, denotes the Ricci scalar
curvature of the $\dst$. The metric $g_{\mu\nu}$ on the Poincar\'e patch of $\dst$ can be read form the line element
\begin{align}\label{metric}
ds^{2}&=\Omega^{2}(\tau)\Big(d\tau^{2}-d\x^{2}\Big), & \Omega(\tau)&=-\frac{1}{H\tau}.
\end{align}
The coordinates conformal time $\tau$ and spatial coordinate $\x$ have ranges
\begin{align}\label{range}
\tau\in\big(-\infty,0\big), && \x\in\mathbb{R},
\end{align}
and cover half of $\dst$ manifold. We consider a uniform electric field background with a constant energy density in the patch (\ref{metric}), which can
be derived from the vector potential
\begin{equation}\label{gauge}
A_{\mu}(\tau)=-\frac{E}{H^{2}\tau}\delta_{\mu}^{1},
\end{equation}
where $E$ is a constant coefficient. Substituting the ingredients (\ref{metric}) and (\ref{gauge}), the Klein-Gordon equation arising from the action (\ref{action}) can be written as
\begin{equation}\label{KG}
\bigg[ \frac{\partial^{2}}{\partial\tau^{2}}-\frac{\partial^{2}}{\partial\x^{2}}-\frac{2i\lambda}{\tau}\frac{\partial}{\partial\x}
+\frac{1}{\tau^{2}}\Big(\frac{1}{4}-\gamma^{2}\Big) \bigg] \varphi(\tau,\x)=0,
\end{equation}
where the definitions of the dimensionless parameters are given by
\begin{align}\label{gamma}
\lambda &=-\frac{eE}{H^{2}}, & \mu &=\frac{m}{H}, & \gamma &=\sqrt{\frac{1}{4}-\lambda^{2}-\mu^{2}-2\xi}.
\end{align}
Since we ultimately wish to compute the expectation value of the energy-momentum tensor in the in-vacuum state, we only require that of the solutions of Eq.~(\ref{KG}) which represent this vacuum sate. Therefor, we impose the boundary condition that in the in region of the manifold as $\tau\rightarrow-\infty$, the mode functions be plane waves of fixed comoving momentum $k$. The normalized positive $U_{k}(x)$, and negative $V_{k}(x)$, frequency mode functions that reduce to the plane wave form in the in region are found to be [see \cite{Garriga:1994bm,Frob:2014zka,Bavarsad:2016cxh} for derivations]
\begin{eqnarray}
U_{k}(x) &=& (2|k|)^{-\frac{1}{2}}e^{\frac{i\pi\kappa}{2}}e^{+ik\x}W_{\kappa,\gamma}\Big(2e^{-\frac{i\pi}{2}}|p|\Big), \label{u}\\
V_{k}(x) &=& (2|k|)^{-\frac{1}{2}}e^{-\frac{i\pi\kappa}{2}}e^{-ik\x}W_{\kappa,\gamma}\Big(2e^{+\frac{i\pi}{2}}|p|\Big), \label{v}
\end{eqnarray}
where the dimensionless physical momentum $p$ and the parameter $\kappa$ are expressed as
\begin{align}\label{kappa}
p &=-\tau k, & \kappa &=-i\lambda r, & r &=\mathrm{sgn}(k).
\end{align}
In Eqs.~(\ref{u}) and (\ref{v}), the factor $W_{\kappa,\gamma}$ denotes the Whittaker function; see, e.g., \cite{NIST}. If the values of the parameters $\kappa,\,\gamma$, and the phase of the variable $z$ satisfy conditions
\begin{align}\label{condit}
\frac{1}{2}\pm\gamma-\kappa \neq 0,-1,-2,\ldots, && \big|\mathrm{ph}(z)\big|<\frac{3}{2}\pi,
\end{align}
then the Whittaker function $W_{\kappa,\gamma}(z)$, with the help of gamma function $\Gamma(z)$, can be represented by a convenient Mellin-Barnes integral
\begin{equation}\label{Mellin}
W_{\kappa,\gamma}(z)=e^{-\frac{z}{2}}\int_{-i\infty}^{+i\infty}\frac{ds}{2\pi i}
\frac{\Gamma\big(\frac{1}{2}+\gamma+s\big) \Gamma\big(\frac{1}{2}-\gamma+s\big)\Gamma\big(-\kappa-s\big)}
{\Gamma\big(\frac{1}{2}+\gamma-\kappa\big)\Gamma\big(\frac{1}{2}-\gamma-\kappa\big)}z^{-s}.
\end{equation}
The contour of integration is a straight line along the imaginary axis in the complex plane $s$ from $-i\infty$ to $+i\infty$ that can be joined by a semicircle at the infinity to sort out the poles of $\Gamma(1/2+\gamma+s)$ and $\Gamma(1/2-\gamma+s)$ from the poles of $\Gamma(-\kappa-s)$.
\par
The mode functions (\ref{u}) and (\ref{v}) satisfy the conserved Wronskian conditions
\begin{equation}\label{wronski}
U_{k}\dot{U}_{k}^{\ast}-U_{k}^{\ast}\dot{U}_{k}=V_{k}^{\ast}\dot{V}_{k}-V_{k}\dot{V}_{k}^{\ast}= i,
\end{equation}
where we use a single dot above a symbol to denote the first conformal time derivative and two dots to denote the seconde conformal time derivative. To quantize the scalar field $\varphi(x)$, we adopt the canonical procedure. Hence, we promote $\varphi(x)$ to operator and expand it in terms of the
compleat set of orthogonal mode functions (\ref{u}) and (\ref{v}) as
\begin{equation}\label{field}
\varphi(x)=\int\frac{dk}{2\pi} \Big[ a_{k}U_{k}(x)+b^{\dag}_{k}V_{k}(x) \Big],
\end{equation}
where the annihilation $a_{k},\,b_{k}$, and creation $a^{\dag}_{k},\,b^{\dag}_{k}$, operators obey the commutation relations
\begin{equation}\label{commut}
\Big[a_{k},a^{\dag}_{k'}\Big]=\Big[b_{k},b^{\dag}_{k'}\Big]=(2\pi)\delta(k-k'),
\end{equation}
with all other commutators equal to zero. Then, we choose the in-vacuum state $|\v\rangle$ to be the state that is annihilated by $a_{k}$ and $b_{k}$ operators
\begin{equation}\label{vacuum}
a_{k}\big| \v \big\rangle=b_{k}\big| \v \big\rangle=0,
\end{equation}
for all values of comoving momentum $k$.
%%%%%%%%%%%%%%%%%%%%%%%%%%%%%%%%%%%%%%%%%%%%%%%%%%%%%%%%%%%%%%%%%%%%%%%%%%%%%%%%%%%%%%%%%%%%%%%%%%%%%%%%%%%%%%%%%%%%%%%%%%%%%%%%%%%%%%%%%%%%%%%%%%%%%%%%%%%%%%%
%%%%%%%%%%%%%%%%%%%%%%%%%%%%%%%%%%%%%%%%%%%%%%%%%%%%%%%%%%%%%%%%%%%%%%%%%%%%%%%%%%%%%%%%%%%%%%%%%%%%%%%%%%%%%%%%%%%%%%%%%%%%%%%%%%%%%%%%%%%%%%%%%%%%%%%%%%%%%%%
\section{\label{sec:compu}computation of the regularized energy-momentum tensor}
We are now ready to compute the expectation value of energy-momentum tensor of the scalar field in the in-vacuum state.
In general, variation of the action $\delta S$ with respect to the inverse metric $\delta g^{\mu\nu}$ defines the energy-momentum tensor as
\begin{equation}\label{general}
T_{\mu\nu}=+\frac{2}{\sqrt{-g}}\frac{\delta S}{\delta g^{\mu\nu}}.
\end{equation}
Vary $g^{\mu\nu}$ in the action (\ref{action}) and use of definition (\ref{general}) along with the Kline-Gordon equation of motion (\ref{KG}), yields the
following symmetric expression for the energy-momentum tensor of the scalar field
\begin{eqnarray}\label{express}
T_{\mu\nu} &=& g_{\mu\nu} \bigg[ \big(4\xi-1\big)g^{\alpha\beta}\Big(\partial_{\alpha}\varphi^{\ast}-ieA_{\alpha}\varphi^{\ast}\Big)
\Big(\partial_{\beta}\varphi+ieA_{\beta}\varphi\Big)-\big(4\xi-1\big)m^{2}\varphi^{\ast}\varphi-4\xi^{2}R\varphi^{\ast}\varphi \bigg] \nn\\
&+& \big(1-2\xi\big)\Big(\partial_{\mu}\varphi^{\ast}\partial_{\nu}\varphi+\partial_{\nu}\varphi^{\ast}\partial_{\mu}\varphi\Big)
+ieA_{\mu}\Big(\varphi\partial_{\nu}\varphi^{\ast}-\varphi^{\ast}\partial_{\nu}\varphi\Big)
+ieA_{\nu}\Big(\varphi\partial_{\mu}\varphi^{\ast}-\varphi^{\ast}\partial_{\mu}\varphi\Big) \nn\\
&+&2e^{2}A_{\mu}A_{\nu}\varphi^{\ast}\varphi+2\xi\Gamma^{\alpha}_{\mu\nu}\Big(\varphi\partial_{\alpha}\varphi^{\ast}
+\varphi^{\ast}\partial_{\alpha}\varphi\Big)-2\xi\Big(\varphi\partial_{\mu}\partial_{\nu}\varphi^{\ast}
+\varphi^{\ast}\partial_{\mu}\partial_{\nu}\varphi\Big),
\end{eqnarray}
where $\Gamma^{\alpha}_{\mu\nu}$ is the Christoffel connection associated with the metric (\ref{metric}) whose nonzero components are
\begin{equation}\label{christo}
\Gamma^{0}_{00}=\Gamma^{0}_{11}=\Gamma^{1}_{01}=\frac{\dot{\Omega}(\tau)}{\Omega(\tau)},
\end{equation}
or related to these by symmetry.
%%%%%%%%%%%%%%%%%%%%%%%%%%%%%%%%%%%%%%%%%%%%%%%%%%%%%%%%%%%%%%%%%%%%%%%%%%%%%%%%%%%%%%%%%%%%%%%%%%%%%%%%%%%%%%%%%%%%%%%%%%%%%%%%%%%%%%%%%%%%%%%%%%%%%%%%%%%%%%%
%%%%%%%%%%%%%%%%%%%%%%%%%%%%%%%%%%%%%%%%%%%%%%%%%%%%%%%%%%%%%%%%%%%%%%%%%%%%%%%%%%%%%%%%%%%%%%%%%%%%%%%%%%%%%%%%%%%%%%%%%%%%%%%%%%%%%%%%%%%%%%%%%%%%%%%%%%%%%%%
\subsection{\label{sec:in}The evaluation of the expectation value in the in-vacuum state}
To evaluate the expectation value of the energy-momentum tensor in the in-vacuum state, we consider $\varphi(x)$ as the scalar field operator and we would
put Eq.~(\ref{field}) into the expression (\ref{express}). Using the relations (\ref{commut}) and (\ref{vacuum}), we then obtain the integral expressions
for the in-vacuum expectation values of the components of the energy-momentum tensor. Changing the integral variable from the comoving  momentum $k$, to
the dimensionless physical momentum $p=-\tau k$, and imposing an ultraviolet cutoff $\Lambda$ on $p$, the measure of integration can be written as
\begin{equation}\label{measure}
\int_{-\infty}^{+\infty}\frac{dk}{(2\pi)} = H\Omega(\tau) \sum_{r=\pm 1}\int_{0}^{\Lambda}\frac{dp}{(2\pi)}.
\end{equation}
Then, the in-vacuum expectation value of the timelike component can be expressed as
\begin{equation}\label{timeli}
\big \langle \v \big| T_{00} \big| \v \big \rangle = \Omega^{2}(\tau)\frac{H^{2}}{(2\pi)}\sum_{r=\pm 1}
\bigg[ \mathcal{I}_{1}-2\lambda r\mathcal{I}_{2}+\Big(\lambda^{2}+\frac{1}{2}\mu^{2}\Big)\mathcal{I}_{3}
+\frac{1}{2}\mathcal{I}_{4}+\mathcal{I}_{5}-2\xi\mathcal{I}_{6}
+\lambda r\mathcal{I}_{7} \bigg],
\end{equation}
where the coefficients $\mathcal{I}_{1},\,\mathcal{I}_{2},\ldots,\mathcal{I}_{7}$ denote the momentum integrals over the Whittaker functions, and are
defined in Eqs.~(\ref{def:I1})-(\ref{def:I7}), respectively. Similarly, the in-vacuum expectation value of the spacelike component is expressed by
\begin{eqnarray}\label{spaceli}
\big \langle \v \big| T_{11} \big| \v \big \rangle&=&\Omega^{2}(\tau)\frac{H^{2}}{(2\pi)}\sum_{r=\pm 1}
\bigg[ \mathcal{I}_{1}-2\lambda r\mathcal{I}_{2}+\Big(\lambda^{2}-\frac{1}{2}\big(1-4\xi\big)\mu^{2}+4\xi^{2}\Big)\mathcal{I}_{3}
+\frac{1}{2}\big(1-4\xi\big)\mathcal{I}_{4} \nn\\
&+&\big(1-4\xi\big)\mathcal{I}_{5}-2\xi\mathcal{I}_{6}+\big(1-4\xi\big)\lambda r\mathcal{I}_{7} \bigg].
\end{eqnarray}
By using Eq.~(\ref{wronski}), it can be verified that the in-vacuum expectation values of the off-diagonal components are equal to
\begin{equation}\label{offdiag}
\big\langle \v \big| T_{01} \big| \v \big\rangle = \big\langle \v \big| T_{10} \big| \v \big\rangle
=\Omega^{2}(\tau)\frac{H^{2}}{\pi}\lambda\Lambda.
\end{equation}
Substituting the expressions (\ref{I1})-(\ref{I7}) into Eqs.~(\ref{timeli}) and (\ref{spaceli}), yields the unregularized in-vacuum expectation values of
the timelike and spacelike components of the energy-momentum tensor, respectively. We find the unregularized timelike component
\begin{eqnarray}\label{unrtime}
\big\langle \v \big| T_{00} \big| \v \big\rangle &=& \Omega^{2}(\tau)\frac{H^{2}}{(2\pi)}
\bigg[ \Lambda^{2}+\mu^{2}\log\big(2\Lambda\big) \nn\\
&-&\xi+\frac{\mu^{2}}{2}+\lambda^{2}
-\frac{\mu^{2}}{2}\Big(1-i\csc(2\pi\gamma)\sinh\big(2\pi\lambda\big)\Big)\psi\Big(\frac{1}{2}+\gamma+i\lambda\Big) \nn\\
&-&\frac{\mu^{2}}{2}\Big(1+i\csc(2\pi\gamma)\sinh(2\pi\lambda)\Big)\psi\Big(\frac{1}{2}-\gamma+i\lambda\Big)
+\lambda\gamma\csc\big(2\pi\gamma\big)\sinh\big(2\pi\lambda\big) \bigg],
\end{eqnarray}
where the notation $\log$ is used to denote the natural logarithm function and $\psi$ denotes the digamma function which is given by the logarithmic
derivative of the gamma function. Also, we find the unregularized spacelike component
\begin{eqnarray}\label{unrspac}
\big\langle \v \big| T_{11} \big| \v \big\rangle &=& \Omega^{2}(\tau)\frac{H^{2}}{(2\pi)}
\bigg[ \Lambda^{2}-\mu^{2}\log\big(2\Lambda\big) \nn\\
&+&\xi+\frac{\mu^{2}}{2}+\lambda^{2}
+\frac{\mu^{2}}{2}\Big(1-i\csc\big(2\pi\gamma\big)\sinh\big(2\pi\lambda\big)\Big)
\psi\Big(\frac{1}{2}+\gamma+i\lambda\Big) \nn\\
&+&\frac{\mu^{2}}{2}\Big(1+i\csc\big(2\pi\gamma\big)\sinh\big(2\pi\lambda\big)\Big)\psi\Big(\frac{1}{2}-\gamma+i\lambda\Big)
+\lambda\gamma\csc\big(2\pi\gamma\big)\sinh\big(2\pi\lambda\big) \bigg].
\end{eqnarray}
We see that the expectation values of the components of the energy-momentum tensor contain ultraviolet divergences.
We will show below that these divergences will be subtracted by the adiabatic counterterms.
%%%%%%%%%%%%%%%%%%%%%%%%%%%%%%%%%%%%%%%%%%%%%%%%%%%%%%%%%%%%%%%%%%%%%%%%%%%%%%%%%%%%%%%%%%%%%%%%%%%%%%%%%%%%%%%%%%%%%%%%%%%%%%%%%%%%%%%%%%%%%%%%%%%%%%%%%%%%%%%
%%%%%%%%%%%%%%%%%%%%%%%%%%%%%%%%%%%%%%%%%%%%%%%%%%%%%%%%%%%%%%%%%%%%%%%%%%%%%%%%%%%%%%%%%%%%%%%%%%%%%%%%%%%%%%%%%%%%%%%%%%%%%%%%%%%%%%%%%%%%%%%%%%%%%%%%%%%%%%%
\subsection{\label{sec:adi}Adiabatic counterterms and regularization of the expectation values}
In order to eliminate the divergent terms of the expressions (\ref{offdiag})-(\ref{unrspac}), we employ the adiabatic regularization procedure. We return
to the Kline-Gordon Eq.~(\ref{KG}) and consider its positive frequency solution as
\begin{equation}\label{adimode}
f(\tau,\x)=e^{+ik\x}h(\tau).
\end{equation}
Then the function $h(\tau)$ satisfies the following field equation
\begin{equation}\label{modeq}
\frac{d^{2}h(\tau)}{d\tau^{2}}+\omega^{2}(\tau)h(\tau)=0,
\end{equation}
and it is convenient to rewrite the conformal time dependent squared frequency as
\begin{equation}\label{omega}
\omega^{2}(\tau)=\omega_{0}^{2}(\tau)+\Delta(\tau),
\end{equation}
where $\omega_{0}(\tau)$ is given by
\begin{equation}\label{omega0}
\omega_{0}(\tau)=+\sqrt{k^{2}+2keA_{1}(\tau)+e^{2}A_{1}^{2}(\tau)+m^{2}\Omega^{2}(\tau)},
\end{equation}
where $A_{1}(\tau)$ is read from Eq.~(\ref{gauge}), and $\Delta(\tau)$ is given by
\begin{equation}\label{delta}
\Delta(\tau)=\frac{2\xi}{\tau^{2}}.
\end{equation}
To adjust the set of the required counterterms, following the usual prescription, we assume that the conformal scale factor $\Omega(\tau)$, and the electromagnetic vector potential $A_{\mu}(\tau)$, to be of zero adiabatic order and the energy-momentum tensor $T_{\mu\nu}$, to be of second adiabatic
order in $\dst$. Therefore, $\omega_{0}(\tau)$ is of zero adiabatic order and $\Delta(\tau)$ which can be rewritten as
\begin{equation}\label{xiterm}
\Delta(\tau)=2\xi\frac{\dot{\Omega}^{2}(\tau)}{\Omega^{2}(\tau)},
\end{equation}
is of seconde adiabatic order. The Klein-Gordon Eq.~(\ref{modeq}) possesses a Wentzel-Kramers-Brillouin (WKB) form solution
\begin{equation}\label{wkb}
h(\tau)=\frac{1}{\sqrt{2\mathcal{W}(\tau)}}\exp\bigg(-i\int^{\tau}d\tau'\mathcal{W}(\tau')\bigg),
\end{equation}
where the function $\mathcal{W}(\tau)$ solves the exact nonlinear second order differential equation
\begin{equation}\label{nonline}
\mathcal{W}^{2}(\tau)=\omega_{0}^{2}(\tau)+\Delta(\tau)-\frac{\ddot{\mathcal{W}}}{2\mathcal{W}}+\frac{3\dot{\mathcal{W}}^{2}}{4\mathcal{W}^{2}}.
\end{equation}
Recall that the set of counterterms which are required to cancel the divergences from the expressions (\ref{offdiag})-(\ref{unrspac}) must be constructed
up to second adiabatic order. It is then necessary to find an adiabatic expansion up to second order for the function $\mathcal{W}$. Thus, we write an appropriate series
\begin{equation}\label{series}
\mathcal{W}(\tau)=\mathcal{W}^{(0)}(\tau)+\mathcal{W}^{(2)}(\tau),
\end{equation}
where the superscripts on the terms indicate their adiabatic orders. The iteration process begins by considering the zeroth adiabatic order. At this step, the adiabatic series (\ref{series}) is truncated to $\mathcal{W}=\mathcal{W}^{(0)}$. Substitution of this ansatz into Eq.~(\ref{nonline}) shows that the derivative terms on the right-hand side of the equation are of second adiabatic order and since the $\Delta$ term is of second adiabatic order too, all these terms vanish. Therefore, we have
\begin{equation}\label{wzero}
\mathcal{W}^{(0)}(\tau)=\omega_{0}(\tau).
\end{equation}
The next iteration is done by substituting the second order adiabatic series (\ref{series}) into Eq.~(\ref{nonline}), using the result (\ref{wzero}) and keeping only terms up to the second order. We then find
\begin{equation}\label{wtwo}
\mathcal{W}^{(2)}(\tau)=\frac{1}{2\omega_{0}}\bigg(\Delta-\frac{\ddot{\omega}_{0}}{2\omega_{0}}+\frac{3\dot{\omega}_{0}^{2}}{4\omega_{0}^{2}}\bigg).
\end{equation}
Thus, the adiabatic expansion of $\mathcal{W}(\tau)$ up to second order is obtained from Eqs.~(\ref{series})-(\ref{wtwo}) as
\begin{equation}\label{wuptwo}
\mathcal{W}(\tau)=\omega_{0}(\tau)
+\frac{1}{2\omega_{0}}\bigg(\Delta-\frac{\ddot{\omega}_{0}}{2\omega_{0}}+\frac{3\dot{\omega}_{0}^{2}}{4\omega_{0}^{2}}\bigg).
\end{equation}
We need also the adiabatic expansion of $\mathcal{W}^{-1}(\tau)$, which up to second order is given by
\begin{equation}\label{winv}
\frac{1}{\mathcal{W}(\tau)}=\frac{1}{\omega_{0}(\tau)}
-\frac{1}{2\omega_{0}^{3}}\bigg(\Delta-\frac{\ddot{\omega}_{0}}{2\omega_{0}}+\frac{3\dot{\omega}_{0}^{2}}{4\omega_{0}^{2}}\bigg).
\end{equation}
Putting together the pieces (\ref{wkb}), (\ref{wuptwo}), and (\ref{winv}) of Eq.~(\ref{adimode}) determines the adiabatic expansion of positive frequency
mode function up to second order. Having these orthogonal adiabatic mode functions $f(x)$, we can perform similar steps which led from Eq.~(\ref{u}) to Eq.~(\ref{vacuum}) and establish the adiabatic expansion of the quantum scalar field operator and the vacuum up to second order. The counterterms are then
obtained by putting the adiabatic expansion of the scalar field operator into Eq.~(\ref{express}) and computing the expectation values of the resulting expressions in the adiabatic vacuum. After these remarks, we find the set of appropriate counterterms as
\begin{eqnarray}
T^{(\mathrm{adi})}_{01} &=& T^{(\mathrm{adi})}_{10}= \Omega^{2}(\tau) \frac{H^{2}}{\pi}\lambda\Lambda, \label{couoffd}\\
T^{(\mathrm{adi})}_{00} &=& \Omega^{2}(\tau) \frac{H^{2}}{(2\pi)} \bigg[ \Lambda^{2}+\mu^{2}\log\big(2\Lambda\big)+\frac{1}{6}-2\xi
+\frac{\mu^{2}}{2}+\lambda^{2}+\frac{\lambda^{2}}{12\mu^{2}}-\mu^{2}\log(\mu) \bigg], \label{coutime} \\
T^{(\mathrm{adi})}_{11} &=& \Omega^{2}(\tau) \frac{H^{2}}{(2\pi)} \bigg[ \Lambda^{2}-\mu^{2}\log\big(2\Lambda\big)-\frac{1}{6}+2\xi
+\frac{\mu^{2}}{2}+\lambda^{2}-\frac{\lambda^{2}}{12\mu^{2}}+\mu^{2}\log(\mu) \bigg]. \label{couspac}
\end{eqnarray}
Subtraction of the counterterms (\ref{couoffd})-(\ref{couspac}) from the unregularized expressions (\ref{offdiag})-(\ref{unrspac}), respectively, yields
the regularized energy-momentum tensor, which is referred to as the induced energy-momentum tensor.
We find that the off-diagonal components of the induced energy-momentum tensor vanish
\begin{equation}\label{emtoffd}
T_{10}=T_{01} = \big \langle \v \big| T_{01} \big| \v\big \rangle - T^{(\mathrm{adi})}_{01}=0.
\end{equation}
The timelike component of the induced energy-momentum tensor is obtained
\begin{eqnarray}\label{emttime}
T_{00} &=& \big\langle\v \big| T_{00} \big|\v\big\rangle - T^{(\mathrm{adi})}_{00} \nn\\
&=& \Omega^{2}(\tau) \frac{H^{2}}{(2\pi)}\bigg[\xi-\frac{1}{6}-\frac{\lambda^{2}}{12\mu^{2}}+\mu^{2}\log(\mu)
-\frac{\mu^{2}}{2}\Big(1-i\csc(2\pi\gamma)\sinh\big(2\pi\lambda\big)\Big)\psi\Big(\frac{1}{2}+\gamma+i\lambda\Big) \nn\\ &-&\frac{\mu^{2}}{2}\Big(1+i\csc(2\pi\gamma)\sinh(2\pi\lambda)\Big)\psi\Big(\frac{1}{2}-\gamma+i\lambda\Big)
+\lambda\gamma\csc\big(2\pi\gamma\big)\sinh\big(2\pi\lambda\big)\bigg].
\end{eqnarray}
Eventually, the spacelike component of the induced energy-momentum tensor is given by
\begin{eqnarray}\label{emtspac}
T_{11} &=& \big\langle\v \big| T_{11} \big|\v\big\rangle - T^{(\mathrm{adi})}_{11} \nn\\
&=&- \Omega^{2}(\tau) \frac{H^{2}}{(2\pi)}\bigg[\xi-\frac{1}{6}-\frac{\lambda^{2}}{12\mu^{2}}+\mu^{2}\log(\mu)
-\frac{\mu^{2}}{2}\Big(1-i\csc(2\pi\gamma)\sinh\big(2\pi\lambda\big)\Big)\psi\Big(\frac{1}{2}+\gamma+i\lambda\Big) \nn\\ &-&\frac{\mu^{2}}{2}\Big(1+i\csc(2\pi\gamma)\sinh(2\pi\lambda)\Big)\psi\Big(\frac{1}{2}-\gamma+i\lambda\Big)
-\lambda\gamma\csc\big(2\pi\gamma\big)\sinh\big(2\pi\lambda\big)\bigg].
\end{eqnarray}
In the case of zero electric field, our result for the induced energy-momentum tensor can be compared to the energy-momentum tensor of a neutral scalar field in $\dst$, which has been derived in Ref.~\cite{Bunch:1978yq} using the covariant point-splitting technique.
If we set $\lambda=0$ in Eqs.~(\ref{emttime}) and (\ref{emtspac}), we then find that the induced energy-momentum tensor can be written as
\begin{equation}\label{emtneut}
T_{\mu\nu} = \frac{H^{2}}{(2\pi)}\bigg[ \xi-\frac{1}{6}+\mu^{2}\log(\mu)
-\frac{1}{2}\mu^{2}\psi\Big(\frac{1}{2}+\gamma\Big) -\frac{1}{2}\mu^{2}\psi\Big(\frac{1}{2}-\gamma\Big) \bigg]g_{\mu\nu}.
\end{equation}
The result (\ref{emtneut}) differs from the corresponding result obtained in \cite{Bunch:1978yq} only by a prefactor of 2, because in \cite{Bunch:1978yq}
a real scalar field has been considered, however here we have considered a complex scalar field $\varphi(x)$, which has two real scalar field components. Thus, the induced energy-momentum tensor, in the zero electric field case, agrees with the energy-momentum tensor of a neutral scalar field obtained earlier.
%%%%%%%%%%%%%%%%%%%%%%%%%%%%%%%%%%%%%%%%%%%%%%%%%%%%%%%%%%%%%%%%%%%%%%%%%%%%%%%%%%%%%%%%%%%%%%%%%%%%%%%%%%%%%%%%%%%%%%%%%%%%%%%%%%%%%%%%%%%%%%%%%%%%%%%%%%%%%%%
%%%%%%%%%%%%%%%%%%%%%%%%%%%%%%%%%%%%%%%%%%%%%%%%%%%%%%%%%%%%%%%%%%%%%%%%%%%%%%%%%%%%%%%%%%%%%%%%%%%%%%%%%%%%%%%%%%%%%%%%%%%%%%%%%%%%%%%%%%%%%%%%%%%%%%%%%%%%%%%
\section{\label{sec:impli}implications of the induced energy-momentum tensor}
In this section we investigate the induced energy-momentum tensor and consider some of its implications.
%%%%%%%%%%%%%%%%%%%%%%%%%%%%%%%%%%%%%%%%%%%%%%%%%%%%%%%%%%%%%%%%%%%%%%%%%%%%%%%%%%%%%%%%%%%%%%%%%%%%%%%%%%%%%%%%%%%%%%%%%%%%%%%%%%%%%%%%%%%%%%%%%%%%%%%%%%%%%%%
%%%%%%%%%%%%%%%%%%%%%%%%%%%%%%%%%%%%%%%%%%%%%%%%%%%%%%%%%%%%%%%%%%%%%%%%%%%%%%%%%%%%%%%%%%%%%%%%%%%%%%%%%%%%%%%%%%%%%%%%%%%%%%%%%%%%%%%%%%%%%%%%%%%%%%%%%%%%%%%
\subsection{\label{sec:analy}Analysis of the induced energy-momentum tensor}
\begin{figure}[t]
\centering
\includegraphics[scale=0.9]{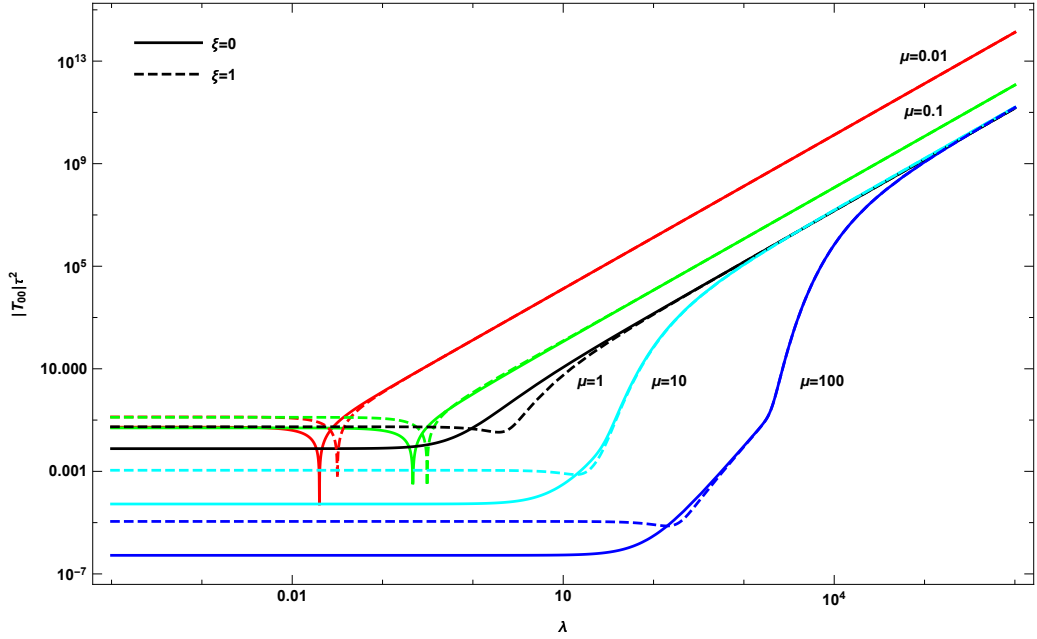}
\caption{The normalized magnitude of the timelike component of the induced energy-momentum tensor $|T_{00}|\tau^{2}$, versus the normalized electric field $\lambda=-eE/H^{2}$, that both scales are logarithmic.
The curves correspond to different values of the mass parameter $\mu=m/H$, and the conformal coupling constant $\xi$.} \label{fig1}
\end{figure}
\begin{figure}[t]
\centering
\includegraphics[scale=0.9]{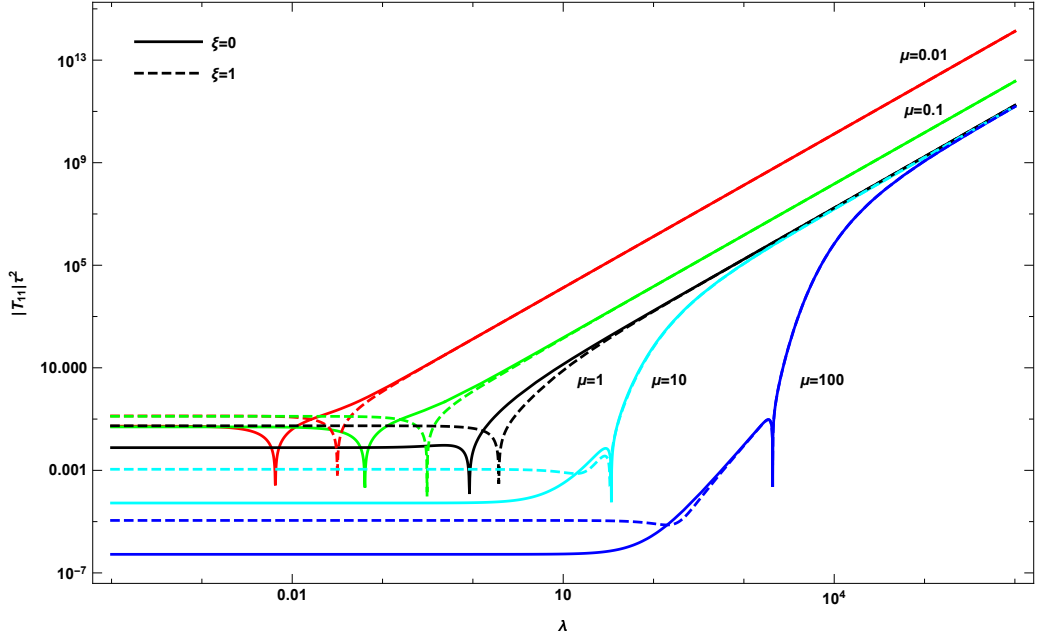}
\caption{The normalized magnitude of the spacelike component of the induced energy-momentum tensor $|T_{11}|\tau^{2}$, versus the normalized electric
field $\lambda=-eE/H^{2}$, that both scales are logarithmic. The curves correspond to different values of the mass parameter $\mu=m/H$, and the conformal coupling constant $\xi$.} \label{fig2}
\end{figure}

We begin our survey of the induced energy-momentum tensor by finding its qualitative behavior. Figures~\ref{fig1} and \ref{fig2} show graphs of the magnitudes of the timelike $|T_{00}|$ [see Eq.~(\ref{emttime})] and spacelike $|T_{11}|$ [see Eq.~(\ref{emtspac})] components of the induced
energy-momentum tensor versus the electric field parameter $\lambda$, respectively. In these figures note especially that the both scales are logarithmic
to cover several orders of magnitude. Several features are clear from these figures. For the cases $\mu\gtrsim 1$, in the strong electric field regime
that the condition $\lambda\gg\max(1,\mu,\xi)$ is valid, $|T_{00}|$ and $|T_{11}|$ are independent of the values of the parameters $\mu$ and $\xi$; hence all the curves asymptotically approach one another at the right end of the figures. Although the expressions (\ref{emttime}) and (\ref{emtspac}) are
rather complicated, they have simple asymptotic forms in the limit $\lambda\rightarrow\infty$, which are given by
\begin{eqnarray}
T_{00} &\simeq& \Omega^{2}(\tau)\frac{H^{2}}{(2\pi)}\lambda^{2}\bigg(1-\frac{1}{12\mu^{2}}\bigg), \label{strongt} \\
T_{11} &\simeq& \Omega^{2}(\tau)\frac{H^{2}}{(2\pi)}\lambda^{2}\bigg(1+\frac{1}{12\mu^{2}}\bigg). \label{strongs}
\end{eqnarray}
The asymptotic behaviors of the curves in Figs.~\ref{fig1} and \ref{fig2}, in the strong electric field regime, are well approximated by Eqs.~(\ref{strongt}) and (\ref{strongs}), respectively. For the cases $\mu<1$, the second terms in both Eqs.~(\ref{strongt}) and (\ref{strongs}), which depend on $\mu$, dominate and as $\mu$ becomes smaller the magnitudes of $T_{00}$ and $T_{11}$ enhance by factor $\mu^{-2}$. While, the first terms in
both Eqs.~(\ref{strongt}) and (\ref{strongs}), which become dominate for the cases $\mu\gtrsim 1$, are independent of the value of $\mu$ and $\xi$.
\par
We see clearly in Figs.~\ref{fig1} and \ref{fig2} the characteristic decrease of the magnitudes of $T_{00}$ and $T_{11}$ at large mass parameter $\mu$,
and the increase at small $\mu$. To find the asymptotic behavior of the induced energy-momentum tensor in the heavy scalar field regime that the condition $\mu\gg\max(1,\lambda,\xi)$ is valid, we can expand expressions (\ref{emttime}) and (\ref{emtspac}) in Taylor series about $\mu=\infty$. We then have
\begin{equation}\label{heavy}
T_{00} \simeq -T_{11} \simeq
\Omega^{2}(\tau)\frac{H^{2}}{\pi} \bigg(\frac{c_{1}}{\mu^{2}}+\frac{c_{2}}{\mu^{4}}+\mathcal{O}\big(\mu^{-6}\big)\bigg),
\end{equation}
where the coefficients $c_{1}$ and $c_{2}$ are given by
\begin{eqnarray}\label{c}
c_{1}&=&\frac{1}{60}-\frac{\xi}{6}+\frac{\xi^{2}}{2}, \nn\\
c_{2}&=& \frac{\lambda^{2}}{20}-\frac{\lambda^{2}\xi}{6}+\frac{2}{315}-\frac{\xi}{15}+\frac{\xi^{2}}{3}-\frac{2\xi^{3}}{3}.
\end{eqnarray}
In the heavy scalar field regime, the approximate expression~(\ref{heavy}) shows that the induced energy-momentum tensor is suppressed as $\mu^{-2}$
instead of an exponentially suppression with a Boltzmann factor $e^{-2\pi\mu}$, which is derived by semiclassical approaches \cite{Bavarsad:2016cxh}.
This behavior have been seen for the in-vacuum expectation value of the energy-momentum tensor of a Dirac field coupled to a uniform electric field in $\dst$ \cite{Botshekananfard:2019zak}. Similar asymptotic behavior occurs in the in-vacuum expectation value of the current of a scalar field in four \cite{Kobayashi:2014zza} and three \cite{Bavarsad:2016cxh} dimensional dS, and also the fermionic induced current in $\dsf$ \cite{Hayashinaka:2016qqn}. Attempts have been made in Refs.~\cite{Banyeres:2018aax,Hayashinaka:2018amz} to address this observation.
\par
Another feature of Figs.~\ref{fig1} and \ref{fig2} is that some of the curves have a singularity. We remark that the graphs have been plotted on the
logarithmic scales; hence the zero values of $T_{00}$ and $T_{11}$ are appeared as extremely sharp decrease in the graph. We stress that the components
of the induced energy-momentum tensor, which are given by Eqs.~(\ref{emtoffd})-(\ref{emtspac}), are continuous and analytic functions of the parameters mass, conformal coupling constant and electric field; as they must \cite{Hollands:2001nf}.
%%%%%%%%%%%%%%%%%%%%%%%%%%%%%%%%%%%%%%%%%%%%%%%%%%%%%%%%%%%%%%%%%%%%%%%%%%%%%%%%%%%%%%%%%%%%%%%%%%%%%%%%%%%%%%%%%%%%%%%%%%%%%%%%%%%%%%%%%%%%%%%%%%%%%%%%%%%%%%%
%%%%%%%%%%%%%%%%%%%%%%%%%%%%%%%%%%%%%%%%%%%%%%%%%%%%%%%%%%%%%%%%%%%%%%%%%%%%%%%%%%%%%%%%%%%%%%%%%%%%%%%%%%%%%%%%%%%%%%%%%%%%%%%%%%%%%%%%%%%%%%%%%%%%%%%%%%%%%%%
\subsection{\label{sec:trace}Trace anomaly}
The trace of the induced energy-momentum tensor $T$, is contracted from the metric (\ref{metric}) and the components (\ref{emtoffd})-(\ref{emtspac}) as
\begin{eqnarray}\label{trace}
T &=& g^{\mu\nu}T_{\mu\nu} \nn\\
&=& \frac{H^{2}}{\pi}\bigg[\xi-\frac{1}{6}-\frac{\lambda^{2}}{12\mu^{2}}+\mu^{2}\log(\mu)
-\frac{\mu^{2}}{2}\Big(1-i\csc(2\pi\gamma)\sinh\big(2\pi\lambda\big)\Big)\psi\Big(\frac{1}{2}+\gamma+i\lambda\Big) \nn\\ &-&\frac{\mu^{2}}{2}\Big(1+i\csc(2\pi\gamma)\sinh(2\pi\lambda)\Big)\psi\Big(\frac{1}{2}-\gamma+i\lambda\Big) \bigg].
\end{eqnarray}
To calculate the trace anomaly, we take the combined lime of Eq.~(\ref{trace}) as $\lambda\rightarrow0$, $\mu\rightarrow0$, and $\xi\rightarrow0$. We then find
\begin{equation}\label{anomaly}
\lim_{\lambda,\,\mu,\,\xi\rightarrow0}T=-\frac{H^{2}}{6\pi}=-\frac{R}{12\pi},
\end{equation}
where in the last step we have used $R=2H^{2}$. The trace anomaly for a real scalar field has been obtained as $(-R)/(24\pi)$ \cite{Duff:1977ay} in a
general two-dimensional spacetime, where $R$ is the Ricci scalar curvature of the spacetime. Here, note in particular that we have regarded a complex
scalar field $\varphi(x)$, which has two real scalar field components. Therefore, the result (\ref{anomaly}) is in agrement with the result obtained in
the literature; see., e.g., \cite{Birrell:1982ix,Parker:2009uva} for a comprehensive review.
%%%%%%%%%%%%%%%%%%%%%%%%%%%%%%%%%%%%%%%%%%%%%%%%%%%%%%%%%%%%%%%%%%%%%%%%%%%%%%%%%%%%%%%%%%%%%%%%%%%%%%%%%%%%%%%%%%%%%%%%%%%%%%%%%%%%%%%%%%%%%%%%%%%%%%%%%%%%%%%
%%%%%%%%%%%%%%%%%%%%%%%%%%%%%%%%%%%%%%%%%%%%%%%%%%%%%%%%%%%%%%%%%%%%%%%%%%%%%%%%%%%%%%%%%%%%%%%%%%%%%%%%%%%%%%%%%%%%%%%%%%%%%%%%%%%%%%%%%%%%%%%%%%%%%%%%%%%%%%%
\subsection{\label{sec:lagra}Effective Lagrangian}
Now that we have obtained the induced energy-momentum tensor, it is possible to return the definition (\ref{general}) and construct the effective action $S_{\e}$ such that its functional derivatives reproduce the expressions (\ref{emtoffd})-(\ref{emtspac}), then we can identify the effective Lagrangian
$\mathcal{L}_{\e}$. We begin by introducing the induced current $J_{\mu}$, which is the regularized in-vacuum expectation value of the current of the
scalar field $\varphi(x)$, whose dynamics is described by the action (\ref{action}). The induced current has been computed in Ref.~\cite{Frob:2014zka},
and is given by
\begin{equation}\label{current}
J_{\mu}=\Omega(\tau)\frac{H}{\pi}e\gamma\csc\big(2\pi\gamma\big)\sinh\big(2\pi\lambda\big)\delta_{\mu}^{1}.
\end{equation}
Thus, the effective electromagnetic potential $A.J$, can be constructed by combining Eqs.~(\ref{gauge}) and (\ref{current}) as
\begin{equation}\label{potenti}
A.J=g^{\mu\nu}A_{\mu}J_{\nu}=\frac{H^{2}}{\pi}\lambda\gamma\csc\big(2\pi\gamma\big)\sinh\big(2\pi\lambda\big).
\end{equation}
To reach our goal of deriving the effective action, it is convenient to rewrite the expressions (\ref{emttime}) and (\ref{emtspac}) in terms of the trace (\ref{trace}) and the effective electromagnetic potential (\ref{potenti}). We then obtain
\begin{eqnarray}
T_{00} &=& \frac{1}{2}\Omega^{2}(\tau) \Big( T+A.J \Big), \label{remttim} \\
T_{11} &=& -\frac{1}{2}\Omega^{2}(\tau)\Big( T-A.J \Big). \label{remtspa}
\end{eqnarray}
Variation of
\begin{equation}\label{seff}
S_{\e}=-\frac{1}{2}\int d^{2}x\sqrt{-g}\Big\{ T+A.J \Big\},
\end{equation}
with respect to the inverse metric $g^{\mu\nu}$ gives
\begin{equation}\label{variati}
\delta S_{\e}=\frac{1}{2}\int d^{2}x\sqrt{-g}\bigg\{\frac{1}{2}\Big(T+A.J\Big)g_{\mu\nu}-A_{\mu}J_{\nu}\bigg\}\delta g^{\mu\nu},
\end{equation}
then definition (\ref{general}), leads to Eqs.~(\ref{emtoffd}), (\ref{remttim}), and (\ref{remtspa}). Therefore, Eq.~(\ref{seff}) is the desired one-loop
effective action of scalar QED in $\dst$, and the corresponding effective Lagrangian reads
\begin{equation}\label{lagrang}
\mathcal{L}_{\e}=-\frac{1}{2}\sqrt{-g}\Big(T+A.J\Big).
\end{equation}
Substitution of Eqs.~(\ref{trace}) and (\ref{potenti}) into Eq.~(\ref{lagrang}) yields the explicit form of the effective Lagrangian as
\begin{eqnarray}\label{leff}
\mathcal{L}_{\e} &=& \sqrt{-g}\Big(\frac{H^{2}}{2\pi}\Big) \bigg[\frac{1}{6}-\xi+\frac{\lambda^{2}}{12\mu^{2}}-\mu^{2}\log(\mu)
+\frac{\mu^{2}}{2}\Big(1-i\csc(2\pi\gamma)\sinh\big(2\pi\lambda\big)\Big)\psi\Big(\frac{1}{2}+\gamma+i\lambda\Big) \nn\\ &+&\frac{\mu^{2}}{2}\Big(1+i\csc(2\pi\gamma)\sinh(2\pi\lambda)\Big)\psi\Big(\frac{1}{2}-\gamma+i\lambda\Big)
-\lambda\gamma\csc\big(2\pi\gamma\big)\sinh\big(2\pi\lambda\big) \bigg].
\end{eqnarray}
The scalar QED effective action in two-dimensional de~Sitter and anti-de~Sitter spacetimes has been obtained in Ref.~\cite{Cai:2014qba}, by employing
the in-out formalism which is introduced by Schwinger and DeWitt; see, e.g., \cite{DeWitt:1975ys} for a review. In the in-out formalism, the effective
action is related to the transition amplitude between in-vacuum and out-vacuum states of the quantum fields; hence it is required to use the Bogoliubov coefficients. In de~Sitter spacetime, in order to have a well-defined out-vacuum state to calculate the Bogoliubov coefficients, it is necessary to adopt the semiclassical approximation.
In the semiclassical regime, the parameters $\lambda,\,\mu$, and $\xi$ are constrained as \cite{Frob:2014zka,Kobayashi:2014zza,Bavarsad:2016cxh}
\begin{equation}\label{semicla}
\lambda^{2}+\mu^{2}+2\xi \gg 1.
\end{equation}
However, the approach that we adopt in this paper involves only the in-vacuum state. Hence, we do not need to consider the out-vacuum sate which in turn requires the condition (\ref{semicla}). Consequently, the effective Lagrangian (\ref{leff}) can be probed in larger domains of the parameters $\lambda,\,\mu$, and $\xi$, compared with those effective Lagrangians which are derived under the semiclassical condition.
%%%%%%%%%%%%%%%%%%%%%%%%%%%%%%%%%%%%%%%%%%%%%%%%%%%%%%%%%%%%%%%%%%%%%%%%%%%%%%%%%%%%%%%%%%%%%%%%%%%%%%%%%%%%%%%%%%%%%%%%%%%%%%%%%%%%%%%%%%%%%%%%%%%%%%%%%%%%%%%
%%%%%%%%%%%%%%%%%%%%%%%%%%%%%%%%%%%%%%%%%%%%%%%%%%%%%%%%%%%%%%%%%%%%%%%%%%%%%%%%%%%%%%%%%%%%%%%%%%%%%%%%%%%%%%%%%%%%%%%%%%%%%%%%%%%%%%%%%%%%%%%%%%%%%%%%%%%%%%%
\section{\label{sec:concl}conclusions}
This paper has investigated the one-loop induced energy-momentum tensor of a complex scalar field in the context of scalar QED in a two-dimensional de~Sitter spacetime. The dynamics of the scalar field is described by the action presented in Eq.~(\ref{action}). We have assumed that the scalar field
propagates in a uniform electric field background in the Poincar\'e patch of $\dst$. The metric of the spacetime can be read from Eq.~(\ref{metric}), and
the electric field background is described by the vector potential (\ref{gauge}). Since, a systematic treatment of ultraviolet divergences in expectation values which are computed in an adiabatic and Hadamard state is relatively simple and straightforward, we calculate the expectation value of
energy-momentum tensor in the in-vacuum state. The results for the expectation values of the energy-momentum tensor components in the in-vacuum state are given by Eqs.~(\ref{offdiag})-(\ref{unrspac}). We have used adiabatic subtraction method to regularize the expectation values, the complete set of the appropriate counterterms is obtained in Eqs.~(\ref{couoffd})-(\ref{couspac}). Then, each of the expressions (\ref{offdiag})-(\ref{unrspac}) is
regularized by subtracting its counterterm. This procedure removes all the ultraviolet divergences and brings us to the induced energy-momentum tensor
whose components are given by Eqs.~(\ref{emtoffd})-(\ref{emtspac}). The components of the induced energy-momentum tensor are continuous and analytic functions of the parameters mass $\mu$, conformal coupling constant $\xi$, and electric field $\lambda$. We showed that, in the zero electric field case,
the induced energy-momentum tensor takes the form (\ref{emtneut}), and agrees with the energy-momentum tensor of a neutral scalar field obtained earlier
in the literature.
\par
We observe that the off-diagonal components of the induced energy-momentum tensor vanish. Figures~\ref{fig1} and \ref{fig2} reveals the behaviors of the magnitudes of timelike $T_{00}$ and spacelike $T_{11}$ components of the induced energy-momentum tensor, respectively. For fixed values of $\mu$ and $\xi$ in the strong electric field regime $\lambda\gg\max(1,\mu,\xi)$, the magnitudes of $T_{00}$ and $T_{11}$ increase significantly with increasing $\lambda$; except in the close neighborhoods of the zero values of $T_{00}$ and $T_{11}$. Recall that in the figures the both scales are logarithmic; hence near the
zero values of $T_{00}$ and $T_{11}$ a singular behavior for the curves is seen. In the strong electric field regime, $T_{00}$ and $T_{11}$ can be well approximated by the expressions (\ref{strongt}) and (\ref{strongs}), respectively. For fixed values of $\lambda$ and $\xi$, the magnitudes of $T_{00}$
and $T_{11}$ decrease with increasing $\mu$. In the heavy scalar field regime $\mu\gg\max(1,\lambda,\xi)$, the approximate expressions for $T_{00}$ and
$T_{11}$ are given by Eq.~(\ref{heavy}).
\par
The trace of the induced energy-momentum tensor has been obtained in Eq.~(\ref{trace}), which yields the trace anomaly (\ref{anomaly}). In the discussion below Eq.~(\ref{anomaly}), we have pointed out that our result for the trace anomaly is in agrement with the trace anomaly of a massless conformally
coupled real scalar field in a general two-dimensional spacetime obtained earlier in the literature.
\par
The major achievement of this research is the derivation of the effective Lagrangian (\ref{leff}) from the induced energy-momentum tensor. More precisely, the expression (\ref{leff}) is the nonperturbative one-loop effective Lagrangian for a scalar field coupled to a uniform electric field background in the Poincar\'e patch of $\dst$. In the derivation of the effective Lagrangian (\ref{leff}), we do not impose any semiclassical condition such as (\ref{semicla}). Consequently, our result for the effective Lagrangian can be examined in larger domains of the parameters $\lambda,\,\mu$, and $\xi$, compared with those effective Lagrangians which are derived by using semiclassical approaches.
%%%%%%%%%%%%%%%%%%%%%%%%%%%%%%%%%%%%%%%%%%%%%%%%%%%%%%%%%%%%%%%%%%%%%%%%%%%%%%%%%%%%%%%%%%%%%%%%%%%%%%%%%%%%%%%%%%%%%%%%%%%%%%%%%%%%%%%%%%%%%%%%%%%%%%%%%%%%%%%
%%%%%%%%%%%%%%%%%%%%%%%%%%%%%%%%%%%%%%%%%%%%%%%%%%%%%%%%%%%%%%%%%%%%%%%%%%%%%%%%%%%%%%%%%%%%%%%%%%%%%%%%%%%%%%%%%%%%%%%%%%%%%%%%%%%%%%%%%%%%%%%%%%%%%%%%%%%%%%%
\begin{acknowledgments}
E.~B.~is supported by the University of Kashan Grant No.~985904/1.
\end{acknowledgments}
%%%%%%%%%%%%%%%%%%%%%%%%%%%%%%%%%%%%%%%%%%%%%%%%%%%%%%%%%%%%%%%%%%%%%%%%%%%%%%%%%%%%%%%%%%%%%%%%%%%%%%%%%%%%%%%%%%%%%%%%%%%%%%%%%%%%%%%%%%%%%%%%%%%%%%%%%%%%%%%
%%%%%%%%%%%%%%%%%%%%%%%%%%%%%%%%%%%%%%%%%%%%%%%%%%%%%%%%%%%%%%%%%%%%%%%%%%%%%%%%%%%%%%%%%%%%%%%%%%%%%%%%%%%%%%%%%%%%%%%%%%%%%%%%%%%%%%%%%%%%%%%%%%%%%%%%%%%%%%%
\appendix*\section{\label{sec:appen}momentum integrals over the Whittaker functions}
In the appendix we present the definitions and explicit values of the coefficients $\mathcal{I}_{1},\mathcal{I}_{2},\ldots,\mathcal{I}_{7}$ which are appeared in Eqs.~(\ref{timeli}) and (\ref{spaceli}). These coefficients are defined as
\begin{flalign}\label{def:I1}
\hspace{1cm}\mathcal{I}_{1} = e^{\pi\lambda r}\int_{0}^{\Lambda}dp p \Big|W_{\kappa,\gamma}\big(-2ip\big)\Big|^{2},&&
\end{flalign}
\begin{flalign}\label{def:I2}
\hspace{1cm}\mathcal{I}_{2} = e^{\pi\lambda r}\int_{0}^{\Lambda}dp \Big|W_{\kappa,\gamma}\big(-2ip\big)\Big|^{2},&&
\end{flalign}
\begin{flalign}\label{def:I3}
\hspace{1cm}\mathcal{I}_{3} = e^{\pi\lambda r}\int_{0}^{\Lambda}\frac{dp}{p}\Big|W_{\kappa,\gamma}\big(-2ip\big)\Big|^{2},&&
\end{flalign}
\begin{flalign}\label{def:I4}
\hspace{1cm}\mathcal{I}_{4} = e^{\pi\lambda r}\int_{0}^{\Lambda}\frac{dp}{p}\Big|W_{1+\kappa,\gamma}\big(-2ip\big)\Big|^{2},&&
\end{flalign}
\begin{flalign}\label{def:I5}
\hspace{1cm}\mathcal{I}_{5} = -e^{\pi\lambda r}\Im\int_{0}^{\Lambda}dp W_{\kappa,\gamma}\big(-2ip\big)W_{1-\kappa,\gamma}\big(2ip\big),&&
\end{flalign}
\begin{flalign}\label{def:I6}
\hspace{1cm}\mathcal{I}_{6} = - e^{\pi\lambda r}\Re\int_{0}^{\Lambda}\frac{dp}{p} W_{\kappa,\gamma}\big(-2ip\big)W_{1-\kappa,\gamma}\big(2ip\big),&&
\end{flalign}
\begin{flalign}\label{def:I7}
\hspace{1cm}\mathcal{I}_{7} = e^{\pi\lambda r}\Im\int_{0}^{\Lambda}\frac{dp}{p} W_{\kappa,\gamma}\big(-2ip\big)W_{1-\kappa,\gamma}\big(2ip\big),&&
\end{flalign}
where the operators $\Im$ and $\Re$ extract the imaginary and real parts of the expressions, respectively. Integrals (\ref{def:I1})-(\ref{def:I7}) are
seen to be of the same type as those momentum integrals which occurred in calculation of the induced current of a scalar field in $\dst$
\cite{Frob:2014zka} and $\dsf$ \cite{Kobayashi:2014zza}. For calculating these integrals, we consider the Mellin-Barnes integral representation of the Whittaker function evaluated on a contour described just below Eq.~(\ref{Mellin}) and accomplish the resulting integrals, as explained in Ref.~\cite{Kobayashi:2014zza} with some routine modifications. We eventually find
\begin{flalign}\label{I1}
\hspace{1cm}\mathcal{I}_{1} &=
\frac{1}{2}\Lambda^{2}+r\lambda\Lambda+\frac{1}{2}\Big(\gamma^{2}+3\lambda^{2}-\frac{1}{4}\Big)\log\big(2\Lambda\big)
+\frac{5}{16}-\frac{7}{4}\lambda^{2}-\frac{1}{4}\gamma^{2}-\frac{3}{2}r\lambda\gamma\csc\big(2\pi\gamma\big)e^{2\pi\lambda r} \nn\\
&-\frac{3}{2}r\lambda\gamma\cot\big(2\pi\gamma\big)-\frac{i}{4}\Big(\gamma^{2}+3\lambda^{2}-\frac{1}{4}\Big)\csc\big(2\pi\gamma\big)
\bigg[ \pi\sin\big(2\pi\gamma\big)+\Big(e^{2\pi\lambda r}+e^{-2\pi i\gamma}\Big) \nn\\
&\times\psi\Big(\frac{1}{2}-\gamma+i\lambda r\Big)-\Big(e^{2\pi\lambda r}+e^{2\pi i\gamma}\Big)\psi\Big(\frac{1}{2}+\gamma+i\lambda r\Big) \bigg],&&
\end{flalign}
and
\begin{flalign}\label{I2}
\hspace{1cm}\mathcal{I}_{2} &=
\Lambda+r\lambda\log\big(2\Lambda\big)-r\lambda-\gamma\cot\big(2\pi\gamma\big)-\gamma\csc\big(2\pi\gamma\big)e^{2\pi\lambda r}
-\frac{i}{2}r\lambda\csc\big(2\pi\gamma\big)\bigg[\pi\sin\big(2\pi\gamma\big) \nn\\
&+\Big(e^{2\pi\lambda r}+e^{-2\pi i\gamma}\Big)\psi\Big(\frac{1}{2}-\gamma+i\lambda r\Big)
-\Big(e^{2\pi\lambda r}+e^{2\pi i\gamma}\Big)\psi\Big(\frac{1}{2}+\gamma+i\lambda r\Big) \bigg].&&
\end{flalign}
Also, integrals (\ref{def:I3})-(\ref{def:I7}) have been computed in Ref.~\cite{Jafari}, by using the procedure explained in \cite{Kobayashi:2014zza}, and
the following results have been obtained
\begin{flalign}\label{I3}
\hspace{1cm}\mathcal{I}_{3} &=
\log\big(2\Lambda\big)-\frac{i}{2}\pi-\frac{i}{2}\csc\big(2\pi\gamma\big)\bigg[
\Big(e^{2\pi\lambda r}+e^{-2\pi i\gamma}\Big)\psi\Big(\frac{1}{2}-\gamma+i\lambda r\Big)-\Big(e^{2\pi\lambda r}+e^{2\pi i\gamma}\Big) \nn\\
&\times\psi\Big(\frac{1}{2}+\gamma+i\lambda r\Big) \bigg],&&
\end{flalign}
\begin{flalign}\label{I4}
\hspace{1cm}\mathcal{I}_{4} =
2\Lambda^{2}-4r\lambda\Lambda+\frac{1}{2}\Big(\frac{1}{4}+\lambda^{2}-\gamma^{2}\Big)+r\lambda\gamma\cot\big(2\pi\gamma\big)
+r\lambda\gamma\csc\big(2\pi\gamma\big)e^{2\pi\lambda r},&&
\end{flalign}
\begin{flalign}\label{I5}
\hspace{1cm}\mathcal{I}_{5} &=
-\Lambda^{2}-\frac{1}{2}\Big(\gamma^{2}+\lambda^{2}-\frac{1}{4}\Big)\log\big(2\Lambda\big)+\frac{1}{4}\pi r\lambda
+\frac{1}{4}\Big(3\lambda^{2}+\gamma^{2}-\frac{5}{4}\Big)+\frac{1}{2}r\lambda\gamma\cot\big(2\pi\gamma\big) \nn\\
&+\frac{1}{2}r\lambda\gamma\csc\big(2\pi\gamma\big)e^{2\pi\lambda r}
+\frac{i}{8}\csc\big(2\pi\gamma\big)\Big(\gamma^{2}+\lambda^{2}-\frac{1}{4}-ir\lambda\Big)
\bigg[\Big(e^{2\pi\lambda r}+e^{-2\pi i\gamma}\Big) \nn\\
&\times\psi\Big(\frac{1}{2}-\gamma+i\lambda r\Big)-\Big(e^{2\pi\lambda r}+e^{2\pi i\gamma}\Big)
\psi\Big(\frac{1}{2}+\gamma+i\lambda r\Big)\bigg]+\frac{i}{8}\csc\big(2\pi\gamma\big) \nn\\
&\times \Big(\gamma^{2}+\lambda^{2}-\frac{1}{4}+ir\lambda\Big)
\bigg[\Big(e^{2\pi\lambda r}+e^{-2\pi i\gamma}\Big)\psi\Big(\frac{1}{2}+\gamma-i\lambda r\Big) \nn\\
&-\Big(e^{2\pi\lambda r}+e^{2\pi i\gamma}\Big)\psi\Big(\frac{1}{2}-\gamma-i\lambda r\Big)\bigg],&&
\end{flalign}
\begin{flalign}\label{I6}
\hspace{1cm}\mathcal{I}_{6} = \frac{1}{2},&&
\end{flalign}
\begin{flalign}\label{I7}
\hspace{1cm}\mathcal{I}_{7} =
2\Lambda-r\lambda-\gamma\cot\big(2\pi\gamma\big)-\gamma\csc\big(2\pi\gamma\big)e^{2\pi\lambda r}.&&
\end{flalign}
%%%%%%%%%%%%%%%%%%%%%%%%%%%%%%%%%%%%%%%%%%%%%%%%%%%%%%%%%%%%%%%%%%%%%%%%%%%%%%%%%%%%%%%%%%%%%%%%%%%%%%%%%%%%%%%%%%%%%%%%%%%%%%%%%%%%%%%%%%%%%%%%%%%%%%%%%%%%%%%
%%%%%%%%%%%%%%%%%%%%%%%%%%%%%%%%%%%%%%%%%%%%%%%%%%%%%%%%%%%%%%%%%%%%%%%%%%%%%%%%%%%%%%%%%%%%%%%%%%%%%%%%%%%%%%%%%%%%%%%%%%%%%%%%%%%%%%%%%%%%%%%%%%%%%%%%%%%%%%%

%%%%%%%%%%%%%%%%%%%%%%%%%%%%%%%%%%%%%%%%%%%%%%%%%%%%%%%%%%%%%%%%%%%%%%%%%%%%%%%%%%%%%%%%%%%%%%%%%%%%%%%%%%%%%%%%%%%%%%%%%%%%%%%%%%%%%%%%%%%%%%%%%%%%%%%%%%%%%%%
%%%%%%%%%%%%%%%%%%%%%%%%%%%%%%%%%%%%%%%%%%%%%%%%%%%%%%%%%%%%%%%%%%%%%%%%%%%%%%%%%%%%%%%%%%%%%%%%%%%%%%%%%%%%%%%%%%%%%%%%%%%%%%%%%%%%%%%%%%%%%%%%%%%%%%%%%%%%%%%
\end{document}